\documentclass[prb,twocolumn,preprintnumbers,amsmath,amssymb,showpacs,%
nofootinbib,floatfix]{revtex4}
\usepackage{graphicx,bm,amsmath,epsfig}

\def\<{\langle}
\def\>{\rangle}

\begin{document}


\def\reff#1{(\ref{#1})}

\def\spose#1{\hbox to 0pt{#1\hss}}
\def\ltapprox{\mathrel{\spose{\lower 3pt\hbox{$\mathchar"218$}}
 \raise 2.0pt\hbox{$\mathchar"13C$}}}
\def\gtapprox{\mathrel{\spose{\lower 3pt\hbox{$\mathchar"218$}}
 \raise 2.0pt\hbox{$\mathchar"13E$}}}

\def\bsigma{\mbox{\protect\boldmath $\sigma$}}
\def\bpi{\mbox{\protect\boldmath $\pi$}}
\def\smfrac#1#2{{\textstyle\frac{#1}{#2}}}
\def\smhalf{ {\smfrac{1}{2}} }

\newcommand{\re}{\mathop{\rm Re}\nolimits}
\newcommand{\im}{\mathop{\rm Im}\nolimits}
\newcommand{\trace}{\mathop{\rm tr}\nolimits}
\newcommand{\fr}{\frac}

\def\Z{{\mathbb Z}}
\def\R{{\mathbb R}}
\def\C{{\mathbb C}}

\title{Comparing different coarse-grained potentials for star polymers}

\author{Roberto Menichetti}
\email{Roberto.Menichetti@roma1.infn.it}
\affiliation{Dipartimento di Fisica, Sapienza Universit\`a di Roma,
P.le Aldo Moro 2, I-00185 Roma, Italy}
\author{Andrea Pelissetto}
\email{andrea.pelissetto@roma1.infn.it}
\affiliation{Dipartimento di Fisica, Sapienza Universit\`a di Roma and 
INFN, Sezione di Roma I, P.le Aldo Moro 2, I-00185 Roma, Italy}

\begin{abstract}
We compare different coarse-grained single-blob models for star polymers.
We find that phenomenological models inspired by the Daoud-Cotton theory
reproduce quite poorly the thermodynamics of these systems, even if the 
potential is assumed to be density dependent, as done in the 
analysis of experimental results. Using the numerically
determined coarse-grained potential,
we also determine the minumum value $f_c$ of 
the functionality of the star polymer for which a fluid-solid
transition occurs. By applying the Hansen-Verlet criterion 
we find $35< f_c\lesssim 40$. This result is confirmed by 
an analysis that uses the modified (reference) hypernetted chain 
method and is qualitatively consistent with previous work. 
\end{abstract}



\maketitle

\section{Introduction}

Soft materials are physical systems of great interest because of their many 
applications. Star polymers, obtained by tethering $f$ polymer chains to a 
central microscopic core, represent a very interesting example. 
Indeed, by changing the 
functionality $f$, one can interpolate between linear chains ($f=1,2$) and 
colloidal spheres, corresponding to $f\gg 1$. Moreover, their particular nature 
is responsible for the appearance of many static and dynamic peculiar features
which are not observed in suspensions of hard-sphere colloids or of linear 
chains.\cite{Grest-96,VFPR-01} 
Full-monomer simulations of star polymers are very 
difficult because of the complexity of the structure of these molecules and of 
the large number of monomers involved. Thus, coarse-grained (CG) models, 
in which each polymer is represented by a single monoatomic molecule
---sometimes it is called blob---have been widely used. 
\cite{Likos-01} To obtain a faithful representation of the star-polymer 
solution thermodynamics one should consider $n$-body interactions among the 
CG molecules.\cite{vonFerber-etal-1,Bolhuis:2001p288,Pelissetto-12}
 A considerable simplification occurs if one only considers the 
dilute regime in which the polymer packing fraction 
$\Phi = 4 \pi R^3_g N/(3 V)$ (here $R_g$ is the zero-density radius of gyration,
while $N$ is the number of star polymers in the volume $V$) is at most of order
1. Indeed, in this case, overlaps are rare and the contributions due to the 
$n$-body potentials are small. Hence, a reasonable approximation is obtained
by only considering the pair potential between the CG molecules, which is defined as
\begin{equation}
    \beta V(r;f) = - \ln \langle e^{-\beta U_{\rm int}} \rangle_r,
\end{equation}
where the average is over all pairs of isolated star polymers 
such that the distance between their centers is $r$; $U_{\rm int}$ is the 
total intermolecular energy. Standard renormalization-group arguments
indicate that, if one considers stars made of $fL$ monomers, the adimensional 
potential $\beta V(r;f)$ is a universal function of $b = r/R_g$ in the limit 
$L\to \infty$. Universality implies that, for any given $f$, the limiting
function ${\beta \cal V}(b;f)$ is independent of the microscopic model. 
The first studies of star polymer solutions at finite density 
based on CG models\cite{LLWAJAR-98} used phenomenological potentials that 
were inspired by the Daoud-Cotton model.\cite{DC-82} They diverged 
logarithmically as $b\to 0$, as predicted theoretically,\cite{WP-86}
and showed an exponential (Yukawa) decrease for $b\to\infty$. 
Such a large-distance behavior is consistent 
with the Daoud-Cotton model predictions and was somewhat confirmed by 
the results of 
Ref.~\onlinecite{RJWFLHRTZ-93}, which found their experimental data to 
be consistent with an effective interaction decaying as $e^{-\alpha r}$. 
Such a behavior was also confirmed---albeit with quite large 
errors---by numerical simulations \cite{JWL-99} of systems with $f \le 50$.
Subsequent numerical work---but
again arms were quite short---did not confirm the Yukawa behavior 
for $f \le 18$.\cite{RF-00} This is not surprising, 
since the Daoud-Cotton model\cite{DC-82}
does not apply for small values of $f$. Hence, the 
phenomenological potential of Ref.~\onlinecite{LLWAJAR-98} cannot be used 
for star polymers with a small number of arms. For these reasons, 
Ref.~\onlinecite{JDLvFL-01} suggested that the potential of 
Ref.~\onlinecite{LLWAJAR-98} should only be used for $f\ge 10$. For smaller
values of $f$,
a second phenomenological potential was proposed.\cite{JDLvFL-01} 
It has the correct logarithmic short-distance behavior and 
shows a Gaussian large-distance decay, in agreement with the 
renormalization-group predictions obtained for linear chains 
($f=2$).\cite{KSB-89} 
A direct numerical determination of the universal
pair potential ${\beta \cal V}(b;f)$ was undertaken by 
Hsu and Grassberger (HG).\cite{HG-04}
By means of a large-scale simulation of an optimal model 
(the Domb-Joyce model\cite{DJ-72} at a specific value of the 
interaction parameter) they obtained 
accurate estimates of the pair potential for several values of $f$ in 
the range $2\le f\le 35$
and provided an accurate parametrization of their results which satisfied all
theoretical constraints. They checked the predicted logarithmic divergence for 
small distances\cite{WP-86} and found a purely Gaussian large-distance decay 
for all values of $f$ investigated. In particular, while at short distances 
their numerical  potential was close to the potential proposed 
in Ref.~\onlinecite{JDLvFL-01}, significant differences were observed at
large distances, as a consequence of the different (Yukawa vs Gaussian) decay.

The phenomenological potentials introduced in 
Refs.~\onlinecite{LLWAJAR-98,JDLvFL-01}
have  been extensively used to study the phase diagram of star polymers,
\cite{WLL-98,WLL-99,LSLWRZ-05} binary star polymer systems,
\cite{ALE-02} mixtures of star polymers and colloids
\cite{DLL-02} and of star and linear polymers,
\cite{CL-10,LCSLZWLR-11} and structural arrest in dense 
star polymer systems.\cite{FSTZLVRDL-03, LVRFTD-04}
Since they are quite different from the exact one derived by
HG,\cite{HG-04} one may question the quantitative and/or qualitative 
validity of the results obtained. It is thus worthwhile to 
repeat these calculations by using all CG models, comparing the 
results obtained by using the phenomenological potentials 
with those obtained by using the accurate expression of the pair potential
obtained by HG,\cite{HG-04} which we take as reference
potential. 

In this paper, we study the thermodynamic behavior of dilute 
star-polymer solutions, by using the CG model based on the 
HG accurate pair potential. We determine
the first virial coefficients, the pressure in the 
dilute regime, and the intermolecular structure factor 
which is needed to compare the theoretical results with the 
experimental ones obtained in scattering experiments. 
These results are then compared with the analogous ones obtained 
by using the phenomenological potentials of 
Refs.~\onlinecite{LLWAJAR-98,JDLvFL-01}. Finally, 
we investigate the phase diagram of star polymer solutions, 
identifying the range of values of $f$ for which 
a liquid-solid transition occurs, again using the HG pair potential.

The paper is organized as follows. In Sec.~\ref{sec2} we introduce
the three different CG models. In Sec.~\ref{sec3} we discuss the 
thermodynamical behavior: first, we compute the second and third virial 
coefficient for each CG model, then the compressibility factor and 
the intermolecular structure factor. In Sec.~\ref{sec3.4} we also 
discuss an extension of the model of Ref.~\onlinecite{JDLvFL-01},
which uses a density dependent corona diameter. Finally, in Sec.~\ref{sec4}
we discuss the phase diagram and in Sec.~\ref{sec5} we present our 
conclusions.

\section{The effective pair potentials: definitions} \label{sec2}

The universal
pair potential ${\beta \cal V}(b;f)$ was determined numerically in 
Ref.~\onlinecite{HG-04} for several values of $f$ between 2 and 35. The final
results were parametrized as 
\begin{equation}
\beta {\cal V}_{HG}(b;f) = {1\over \tau_f} 
  \ln \left[ e^{\tau_f V_{WP}(b) - d_f b^2} + 
             e^{\tau_f V_G(b)} \right],
\label{potential-HG}
\end{equation}
\begin{equation}
V_{WP}(b)= b_f \ln (a_f/b) \qquad \qquad
V_G(b) = c_f e^{-d_f b^2}.
\end{equation}
The potential  depends on five constants $a_f$, $b_f$, $c_f$, $d_f$, $\tau_f$ 
which are reported in Ref.~\onlinecite{HG-04}. For $b\to 0$, 
${\cal V}(b;f)$ diverges \cite{WP-86,vonFerber-etal-2} 
as $\ln 1/b$ with a prefactor that 
can be expressed in terms of the partition-function exponents $\gamma_f$,
which are known with good precision.\cite{HNG-04} 
Parametrization (\ref{potential-HG}) satisfies this property, 
${\beta \cal V}_{HG}(b;f)\approx b_f\ln (a_f/b)$, the coefficient
$b_f$ being equal to the theoretically predicted value. For $b\to \infty$
the potential behaves as $c_f e^{-d_f b^2}$, where $d_f$ varies 
between 0.405 ($f=2$) and 0.68 ($f=35$).

Numerical studies of the properties of star polymers have often relied on 
phenomenological expressions for the pair potential. In 
Ref.~\onlinecite{LLWAJAR-98} the following potential was proposed: 
\begin{eqnarray}
\beta {\cal V}_1(R;f) =
     {5 f^{3/2}\over 18} 
     \left(-\ln R + K_f\right)
    && \qquad R \le 1, \nonumber \\
\qquad = 
     {5 f^{3/2}\over 18} 
   {K_fe^{-\sqrt{f} (R - 1)/2}\over R}
     && \qquad R > 1, 
\label{potential-1} 
\end{eqnarray}
where $K_f = 1/(1 + \sqrt{f}/2)$,
$R = r/\sigma$, and $\sigma$ is the so-called corona diameter. For $R\to 0$
the potential shows the expected logarithmic behavior
$\hat{b}_f \ln 1/R$ with $\hat{b}_f = 5 f^{3/2}/18$. 
The coefficient $\hat{b}_f$ can be compared with the theoretical 
result $b_f$ obtained by using the accurate estimates of the partition-function
exponents $\gamma_f$.\cite{HNG-04} We obtain 
${b}_f = 2.42(1), 9.90(3), 57.3(6)$ for $f = 4,10,30$, to be compared with
$\hat{b}_f = 2.22, 8.78, 45.6$ for the same values of $f$.
Differences increase with $f$ and range from 8\% for $f=4$ to 20\% for $f=30$.
They are, however, expected to be largely irrelevant for the 
thermodynamics in the dilute regime, 
in which overlaps are rare. For $R\to \infty$ 
the potential behaves as $e^{-\sqrt{f}R/2}/R$,
which is quite different from the behavior observed in numerical simulations
(at least for $f \le 35$), see Eq.~(\ref{potential-HG}).

Potential (\ref{potential-1}) is expected to be reliable only for large values 
of $f$, for $f > 10$, say, i.e.
in the regime to which the Daoud-Cotton model\cite{DC-82} applies. 
For small values of $f$ a different potential was postulated:
\cite{JDLvFL-01}
\begin{eqnarray}
\beta {\cal V}_2(R;f) =
     {5 f^{3/2}\over 18} 
     \left(-\ln R + \displaystyle{1\over 2\tau^2}
    \right) && \qquad R \le 1, \nonumber \\
\qquad =
     {5 f^{3/2}\over 36\tau^2} e^{-\tau^2 (R^2 - 1)}
     && \qquad R > 1, 
\label{potential-2} 
\end{eqnarray}
where again $R = r/\sigma$. The 
adimensional parameter $\tau$ determines the large-$r$ behavior of 
the potential and was determined only for $f = 2$ and $f = 5$: 
$\tau = 1.03$ and $\tau = 1.12$ in the two cases. 

In the following we will compare the predictions of the three CG models defined above: 
model MHG based on the exact pair potential (\ref{potential-HG}) and models M1 and M2
based on potentials (\ref{potential-1}) and (\ref{potential-2}), respectively.

\section{Results: Structure and thermodynamics} \label{sec3}

\subsection{Zero-density results}  \label{sec3.1}

\begin{table}
\squeezetable
\caption{In the second, third, and fourth column we report
virial coefficient ratios computed by using potential
${\cal V}_{HG}(b;f)$ (HG). In the last three columns we report literature 
values obtained from full-monomer (FM) simulations:
$a$ refers to Ref.~\onlinecite{CMP-06},
$b$ to Ref.~\onlinecite{CMP-08},
$c$ to Ref.~\onlinecite{LK-02}, 
$d$ to Ref.~\onlinecite{Randisi-13},
$e$ to Ref.~\onlinecite{WFST-09}. 
For a review of older estimates of $A_2$, see Ref.~\onlinecite{DRF-90}.
}
\label{table-virial-HG}
\begin{tabular}{ccccccc}
\hline\hline
$f$  &$A_2$(HG)&$A_3$(HG)&$g$(HG)&  $A_2$(FM)&  $A_3$(FM)& $g$(FM)  \\
\hline
2    &  5.51 & 4.97 & 0.164 &5.500(3)$^a$& 9.80(2)$^a$ & 0.324(1)$^a$ \\
4    & 10.04 & 31.8 & 0.316 &9.979(9)$^b$& 39.56(16)$^b$ & 0.397(2)$^b$ \\
5    & 12.25 & 54.5 & 0.363 &           &           &       \\
6    & 14.66 & 85.6 & 0.398 &14.174(16)$^b$& 90.1(0.4)$^b$& 0.449(2)$^b$ \\
10   & 20.62 & 204  & 0.479 &           &           &              \\
12   & 23.97 & 288  & 0.501 &23.5(2)$^c$&           &              \\
18   & 29.60 & 473  & 0.540 &29.6(2)$^c$&           &  0.547(5)$^d$ \\
     &  &  &  &&            &  0.58$^e$\\
24   & 36.06 & 721  & 0.557 &34.21(10)$^d$&661(6)$^d$& 0.564(4)$^d$  \\
30   & 38.32 & 831  & 0.565 &37.65(6)$^d$& 813(7)$^d$& 0.574(6)$^d$  \\
\hline\hline
\end{tabular}
\end{table}

Knowledge of the pair potential ${\cal V}(b;f)$ allows us to compute 
the universal combination $A_2 = B_2/R_g^3$, 
where the second virial coefficient 
$B_2$ is defined by the expansion of the (osmotic) pressure $\Pi$,  
\begin{equation}
{\Pi\over k_BT\rho} = 1 + B_2 \rho + B_3 \rho^2 + O(\rho^3),
\end{equation}
in powers of the concentration $\rho = N/V$. 
Indeed, $A_2$ is related to the pair potential by the exact relation
\begin{equation}
A_2(f) = 2 \pi \int_0^\infty d b\, b^2 \left(
    1 - e^{-{\beta \cal V}(b;f)} \right).
\end{equation}
In Table \ref{table-virial-HG} 
we report the estimates of $A_2$ for several values of $f$ obtained by using 
potential (\ref{potential-HG}). If we compare these results with those obtained in
the literature from different full-monomer simulations [column $A_2$(FM)] 
we observe reasonable agreement (the largest deviation, 6.5\%,
is observed for $f=24$), confirming the adequacy of parametrization 
(\ref{potential-HG}) for all values of $f$. 

Using model MHG we can also compute 
the universal combinations $A_3 = B_3 R_g^{-6}$ and 
$g = B_3/B_2^2 = A_3/A_2^2$, involving the third virial coefficient $B_3$. 
Since three-body interactions are neglected in the CG model, 
these estimates differ
from those that would be obtained in the exact, full-monomer  polymer model. 
Therefore, the observed discrepancies 
give us quantitative indications of the role of the neglected 
many-body interactions in CG single-blob star-polymer models. 
The results reported in Table \ref{table-virial-HG} 
show that many-body forces apparently become less relevant as $f$ increases. 
For instance, if we consider the relative deviation 
$\Delta g = 1 - g({\rm HG})/g({\rm FM})$ for the $g$
parameter, we find a very large discrepancy for $f=2$, 
$\Delta g \approx 50$\%, 
but only $\Delta g \approx 11$\% for $f = 6$. 
For $f \gtrsim 18$, the CG model apparently reproduces the full-monomer
results.  This is particularly encouraging 
since it implies that CG models provide increasingly better approximations as $f$ 
increases, i.e. exactly in the regime in which full-monomer simulations become 
unfeasible. 

\begin{table}
\caption{Estimates of $\sigma/R_g$, $A_2$, and $A_3$ for models 
M1 and M2.
}
\label{table-virial-M1-M2}
\begin{tabular}{ccccccc}
\hline\hline
  & \multicolumn{3}{c}{M1} & \multicolumn{3}{c}{M2} \\
$f$  &$\sigma/R_g$(M1)&$A_3$(M1)&$g$(M1)&$\sigma/R_g$(M2)&  $A_3$(M2)& $g$(M2)  \\
\hline
2    &  0.807 & 2.01 & 0.067 &  1.333 & 6.35 & 0.209 \\
4    &  0.894 & 16.4 & 0.163 &        &      &  \\
5    &  0.933 & 30.6 & 0.204 &  1.349 & 61.9 & 0.412  \\
6    &  0.974 & 50.9 & 0.237 &        &      &  \\
10   &  1.062 & 141  & 0.332 &        &      &  \\
18   &  1.195 & 376  & 0.429 &        &      &  \\
30   &  1.328 & 722  & 0.492 &        &      &  \\
\hline\hline
\end{tabular}
\end{table}

\begin{figure}[tb]
\begin{tabular}{c}
\epsfig{file=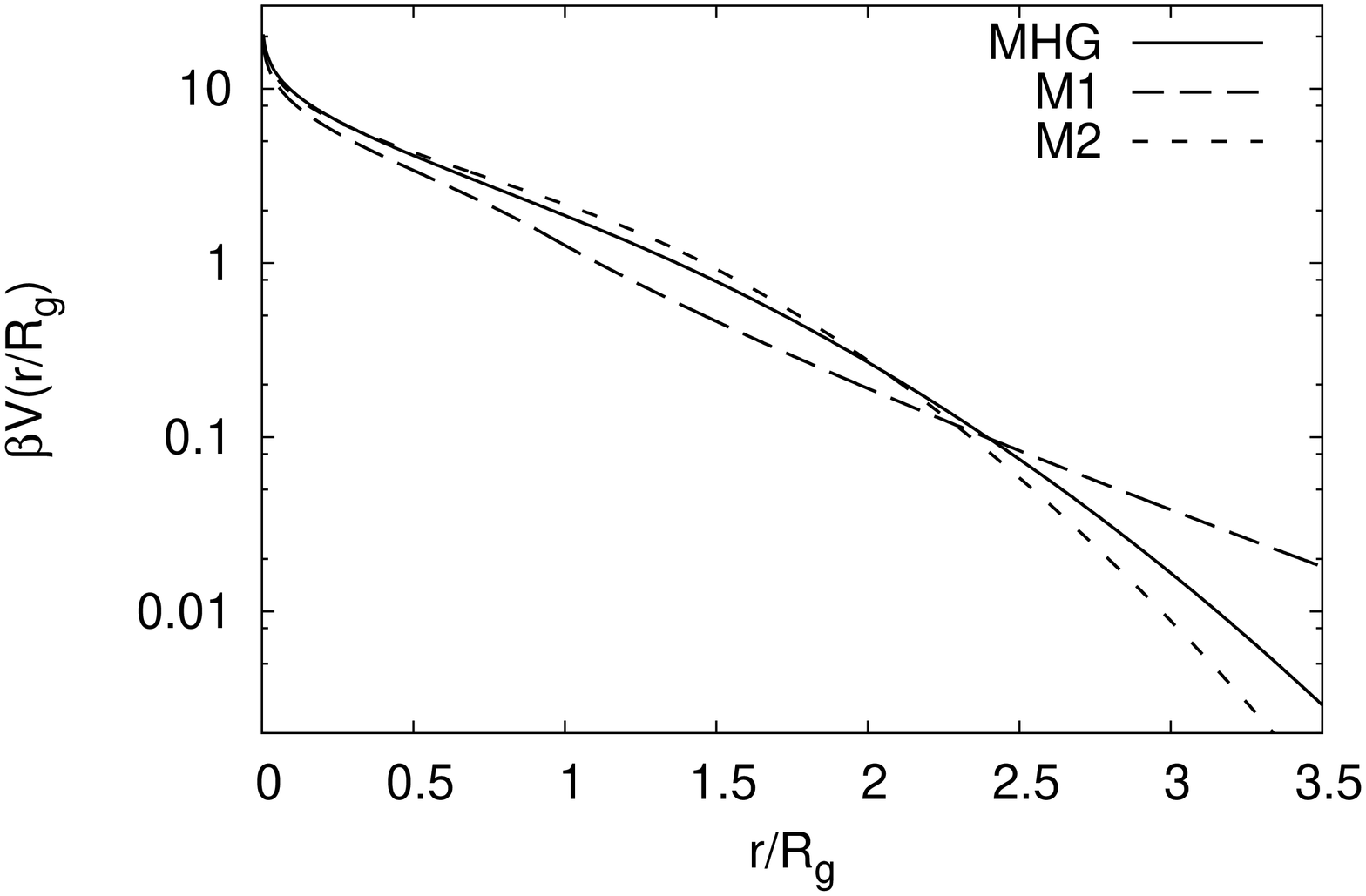,angle=-90,width=8truecm} \hspace{0.5truecm} \\
\epsfig{file=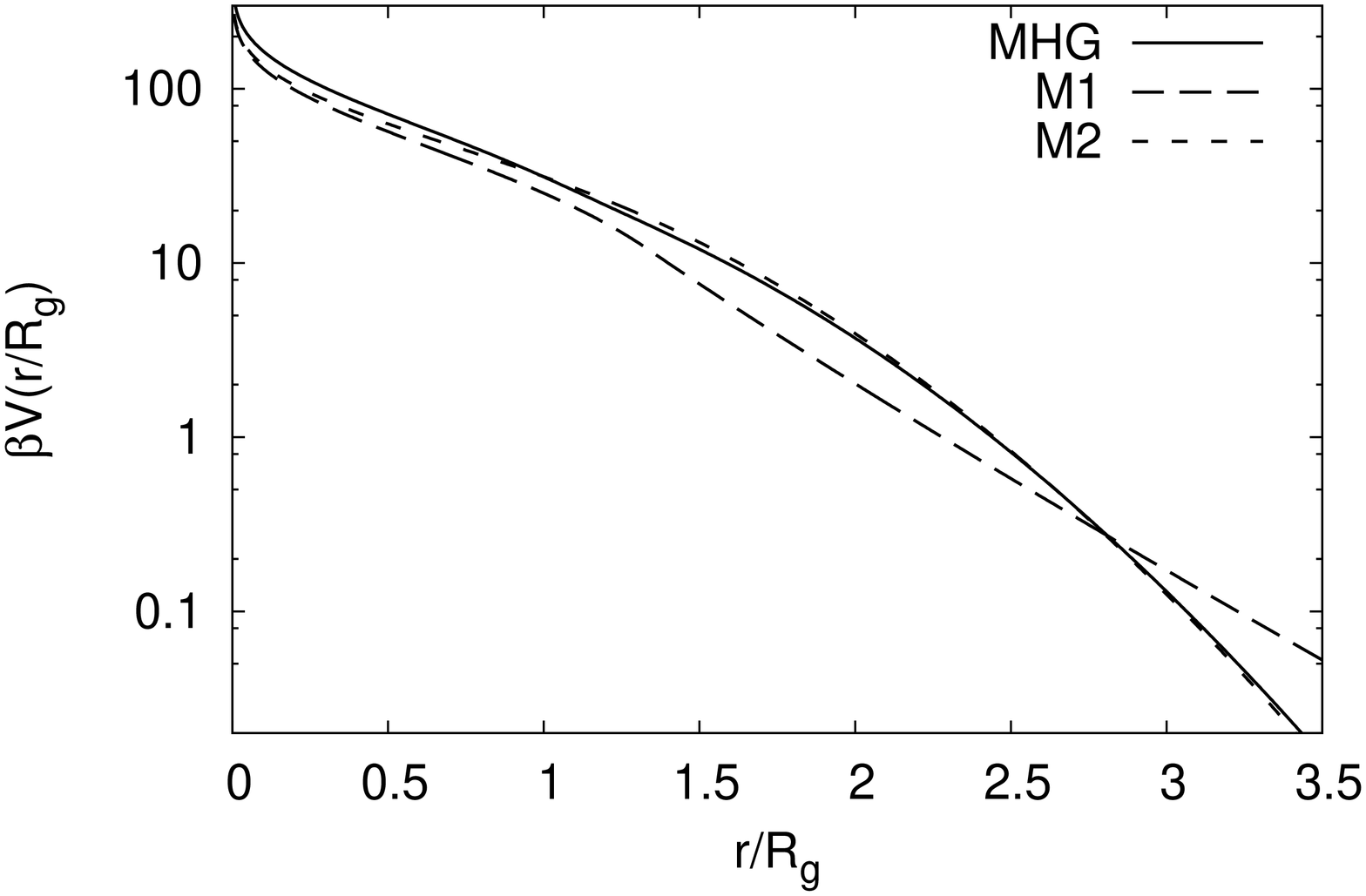,angle=-90,width=8truecm} \hspace{0.5truecm} 
\end{tabular}
\caption{Comparison of the CG potentials of models MHG, M1, and M2: 
(top) $f=5$; (bottom) $f=30$. 
For $f=30$ we fix $\tau = 1$ in model M2 (see Sec.~\protect\ref{sec3.3}).
}
\label{figpotenziali}
\end{figure}

Let us now perform the same analysis for models M1 and M2. Since 
the potentials ${\cal V}_1(r;f)$ and ${\cal V}_2(r;f)$ depend on the corona 
diameter $\sigma$, to obtain quantitative predictions we must determine the 
ratio $\sigma/R_g$. As already suggested in 
Ref.~\onlinecite{DLL-02},
 we fix $\sigma/R_g$ so that all potentials give the correct
result for the second-virial combination $A_2$. This guarantees that models 
M1 and M2 have the correct thermodynamic behavior as $\Phi\to 0$.
Using the estimates of $A_2$ reported in Table~\ref{table-virial-HG} [column
$A_2$(HG)] we obtain the estimates of $\sigma/R_g$ 
reported in Table~\ref{table-virial-M1-M2}.
For $f = 18$, Ref.~\onlinecite{LLWAJAR-98} obtained $\sigma/R_g \approx 1.26$ 
from the analysis
of the experimental data, while numerical simulations \cite{JWL-99} indicate 
that $\sigma/R_g \approx 1.3$ is a good approximation for model M1 at least for $f$ large.
Our numerical results are fully consistent with this approximation. 
As long as $f$ is larger 
than 10, the expected range of validity of model M1, 
$\sigma/R_g \approx 1.3$  holds with a relative error of at most
20\%. Such an approximation 
also holds for model M2 for the two values of $f$ we consider. 
In the following, we consider models M1 and M2 using the values 
of $\sigma/R_g$ 
reported in Table~\ref{table-virial-M1-M2}. Hence, by construction, all models 
(M1, M2, and MHG) have the same thermodynamic behavior for $\Phi\to 0$. 
In Fig.~\ref{figpotenziali} we compare the different potentials for 
$f=5$ and $f=30$. For $f=5$ potentials M2 and MHG are very close in the whole
interesting range $b = r/R_g \lesssim 2.5$. For larger values, 
potential M2 decreases slightly faster:
$\beta {\cal V}_{HG}$ decays as $1.76 e^{-0.53 b^2}$, while 
$\beta {\cal V}_{M2}$ decays as $4.34 e^{-0.69 b^2}$.
Significant differences are instead observed for potential M1, both for 
small values of $b$ --- it underestimates $\beta {\cal V}_{HG}$---and 
for large values of $b$, where it decays slower. Similar discrepancies
are observed for $f=30$.

Let us now compare the 
third-virial coefficient combinations $A_3$ and $g$, 
see Tables~\ref{table-virial-HG} and \ref{table-virial-M1-M2}.
For model M1, discrepancies 
are quite large for $f$ small, confirming the inadequacy of the 
parametrization for these values of $f$. 
For $f > 10$ discrepances are smaller, but still not 
negligible. For $f=18$ $g$(M1) differs from $g$(HG) and $g$(FM) by 21\%.
For $f=30$ the discrepancy decreases to 13\%. 
For $f \le 10$ model M2 should be used. Also in this case, we observe significant discrepances
from the results obtained by using the MHG model, 
but, at least for $f=2$, model M2 appears to 
provide a better approximation to the full-monomer results. 

\subsection{Integral-equation methods} \label{sec3.2}

\begin{figure}[tb]
\begin{tabular}{c}
\epsfig{file=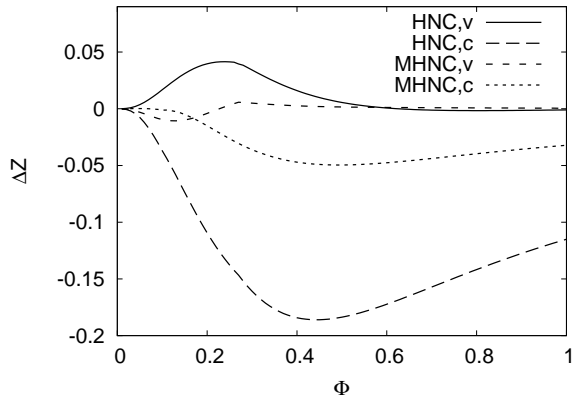,angle=-90,width=8truecm} \hspace{0.5truecm} 
\vspace{0.5truecm}
\end{tabular}
\caption{Relative deviations $\Delta Z_{HNC} = Z_{HNC}/Z_{RY}-1$ and
$\Delta Z_{MHNC} = Z_{MHNC}/Z_{RY}-1$ for $f=30$ and model MHG. 
The virial route (HNC,v and MHNC,v) and the 
compressibility route (HNC,c and MHNC,c) have been used.
}
\label{inteq-comparison}
\end{figure}

In order to study the finite-density behavior we used 
integral-equation methods.\cite{HansenMcDonald} 
As usual in these approaches we considered the pair distribution function
$g({\bf r})$, the corresponding correlation function 
$h({\bf r}) = g({\bf r})-1$,
and the direct correlation function $c({\bf r})$ defined by the 
Ornstein-Zernike relation\cite{HansenMcDonald}
\begin{equation}
h({\bf r}) = c({\bf r}) + \rho \int d^3{\bf s}\, 
     c({\bf s}) h({\bf r}-{\bf s}).
\end{equation}
This equation must be supplemented by a closure relation.
Three different closures were used to check
the accuracy of the results. 
We first used the hypernetted chain (HNC) equation \cite{HansenMcDonald}
\begin{equation}
   g({\bf r}) = e^{-\beta V({\bf r}) + h({\bf r}) - c({\bf r})},
\end{equation}
which is known to be quite accurate for soft interactions. 
In our case potentials diverge as $r\to 0$, hence a better 
approximation should be provided by the 
Rogers-Young (RY) closure:\cite{RY-84}
\begin{equation}
g({\bf r}) = e^{-\beta V({\bf r})} \left\{ 
   1 + {1\over f(r)} \left[ e^{(h({\bf r}) - c({\bf r}))f(r)} - 1\right]
    \right\},
\end{equation}
with 
\begin{equation}
   f(r) = 1 - e^{-\alpha r}.
\end{equation}
The consistency parameter $\alpha$ was redetermined at each density by
requiring the equality of the compressibility computed by using $g({\bf r})$
(compressibility route) and that computed by using the virial pressure. 
This was done iteratively until the relative difference between the two
quantities was less than 0.1\%. We found that 
$\alpha R_g$ increases with $\Phi$, varying between 
0.5  and 2-2.5 as $\Phi$ increases from 0.3 to 1. 

For $f\ge 30$ --- in this case the potential has a quite hard core ---
we also used the reference or modified
HNC (MHNC) method.\cite{RA-79,HansenMcDonald}
The closure relation is written as 
\begin{equation}
g({\bf r}) = e^{-\beta V({\bf r}) + h({\bf r}) - c({\bf r}) + E({\bf r})},
\end{equation}
where $E({\bf r})$ is the bridge function. For $E({\bf r})$ we 
used the bridge function of a system of hard spheres of diameter $d$ 
at the same density (it can be computed quite 
precisely by using the results reported in Refs.~\onlinecite{VW-72,HG-75}). 
The diameter $d$, or equivalently the hard-sphere packing fraction 
$\eta_{HS} = \pi d^3 \rho/6$, was determined by using the Lado 
criterion,\cite{Lado-82} 
which is a way to implement thermodynamic consistency between the 
virial and the energy route:
\begin{equation}
\int d^3{\bf r}\, [g({\bf r}) - g_{HS}({\bf r};\eta_{HS})] 
    {\partial E({\bf r};\eta_{HS})\over \partial \eta_{HS}} = 0. 
\end{equation}
To check the accuracy of the results we computed the 
compressibility factor 
\begin{equation}
    Z= {P\over k_BT\rho}
\end{equation}
for model MHG by using the HNC and the RY closure; for $f\ge 30$ also the 
MHNC closure was used. As an example,
in Fig.~\ref{inteq-comparison} we show $\Delta Z_{HNC} = Z_{HNC}/Z_{RY} - 1$
and $\Delta Z_{MHNC} = Z_{MHNC}/Z_{RY} - 1$, where $Z_{HNC}$, $Z_{RY}$, and 
$Z_{MHNC}$ are the compressibility factors computed by using the three
different methods. Here we take $f=30$ and use potential MHG. 
For the HNC closure, we observe a significant difference between the pressure
computed by using the virial and the compressibility route, which however
decreases as $\Phi$ increases. For the MHNC closure, results are much 
more consistent, the difference being at most 5\%, again decreasing 
as $\Phi$ becomes large. The HNC and MHNC virial pressure differ only slightly
from the RY result. If the compressibility route is used, 
differences are larger. They however decrease as the density
increases. The somewhat large difference between the two thermodynamic routes 
observed for the HNC method is due 
to the somewhat hard core of the potential, that diverges as $r\to 0$. 
It is smaller for smaller values of $f$: for $f = 6$ and 18, $Z_{HNC,c}$ and
$Z_{HNC,v}$ differ at most by 2.2\% and 14\%, respectively. In all cases
the RY closure
appears to be reliable, with an error that is probably of the order
of a few percent and that decreases as $\Phi$ increases.

The results we will present in the following have been obtained by using 
the RY closure. We will use again the MHNC closure in Sec.~\ref{sec4}, where 
we will discuss the fluid-solid transition in star-polymer solutions.

\subsection{Comparing the potentials at finite density} \label{sec3.3}

\begin{figure}[tb]
\begin{tabular}{c}
\epsfig{file=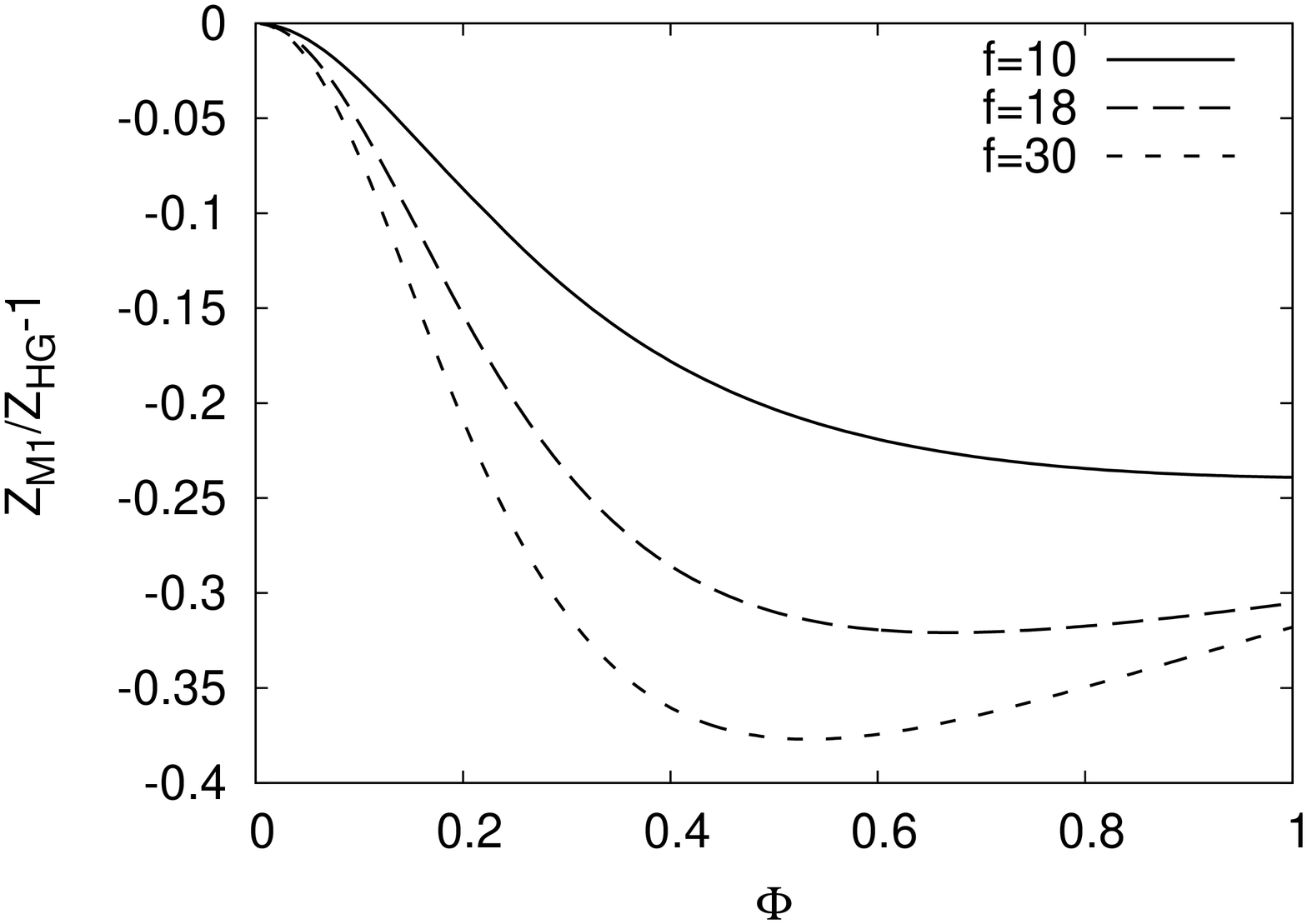,angle=-90,width=7truecm} \hspace{0.5truecm} \\
\epsfig{file=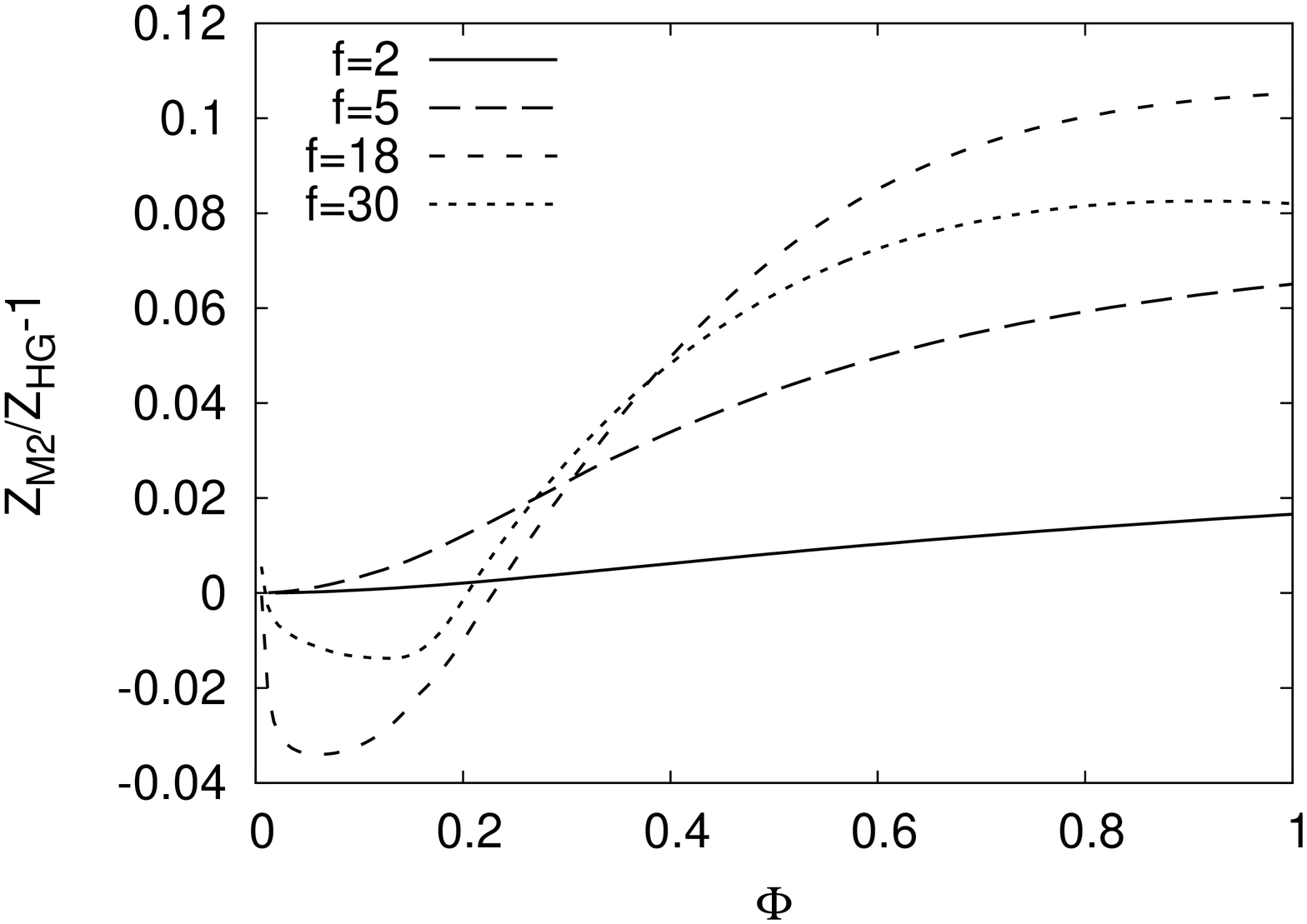,angle=-90,width=7truecm} \hspace{0.5truecm} \\
\end{tabular}
\caption{Relative deviation $\Delta Z = Z/Z_{HG}-1$ 
for model M1 ($f=10,18,30$) (top) and model M2 ($f=2,5,18,30$) (bottom) as 
a function of the polymer volume fraction $\Phi$. $Z_{HG}$ is the 
compressibility factor obtained by using potential
(\protect\ref{potential-HG}).
}
\label{figDeltaZ}
\end{figure}

Let us now discuss the behavior of the different models at finite density in
the dilute regime $\Phi\lesssim 1$. 
Let us first consider the compressibility factor $Z$. In Fig.~\ref{figDeltaZ}
we report the relative deviations $\Delta Z = Z/Z_{HG}-1$, 
where $Z$ is computed in models M1 and M2, and $Z_{HG}$ is computed in 
model MHG. As should be expected on the basis of 
the results for $A_3$ and $g$, model
M1 understimates the true CG compressibility factor $Z_{HG}$.
For $\Phi = 1$, we find deviations $\Delta Z = -24$\%, $-$31\%, $-$32\% for 
$f=10,18,30$, which are quite significant. Model M2 instead overestimates 
$Z_{HG}$. Differences are, however, significantly smaller: 
for $\Phi = 1$, we find $\Delta Z = 2$\%, 6\% for $f=2,5$. 

Since model M2 has the observed Gaussian decay and appears to be relatively
accurate, we tried to check whether it is possible to extend it to 
other values of $f$, beside $f = 2, 5$. 
For this purpose we should fix both $\sigma/R_g$ and $\tau$. 
If we require $g({\rm M2})\approx g({\rm HG})$ beside 
$A_2({\rm M2})\approx A_2({\rm HG})$, we obtain $\tau \approx 1$ 
for all values of $f$ in the range $10\le f \le 35$. 
The ratio $\sigma/R_g$ is always consistent with 1.2. More precisely,
for $\tau = 1$, we have $\sigma/R_g = 1.203,1.202$ for $f=18,30$, 
respectively. The corresponding
M2 potential has 
a large-distance behavior which is consistent with that of potential
(\ref{potential-HG}). Indeed, we have 
$\beta {\cal V}_2(R;f) \sim e^{-0.69 b^2}$ for large $b=r/R_g$ to be 
compared with $\beta {\cal V}_{HG}(b;f) \sim e^{-d_f b^2}$, 
$d_f = 0.65, 0.68$ for $f = 18,30$, respectively. 
Potential $\beta {\cal V}_2(R;f)$ is compared with $\beta {\cal V}_{HG}(b;f)$
in Fig.~\ref{figpotenziali}. On the scale of the figure, no significant 
differences are observed, confirming that model M2 is a better approximation 
to the star-polymer CG potential than model M1.
In Fig.~\ref{figDeltaZ} we also report $\Delta Z$ for model M2 with 
$\tau = 1$ and $f=18,30$. Again model M2 overestimates $Z_{HG}$, 
but differences are only of order 10\% at most.

\begin{figure*}[tb]
\begin{tabular}{cc}
\epsfig{file=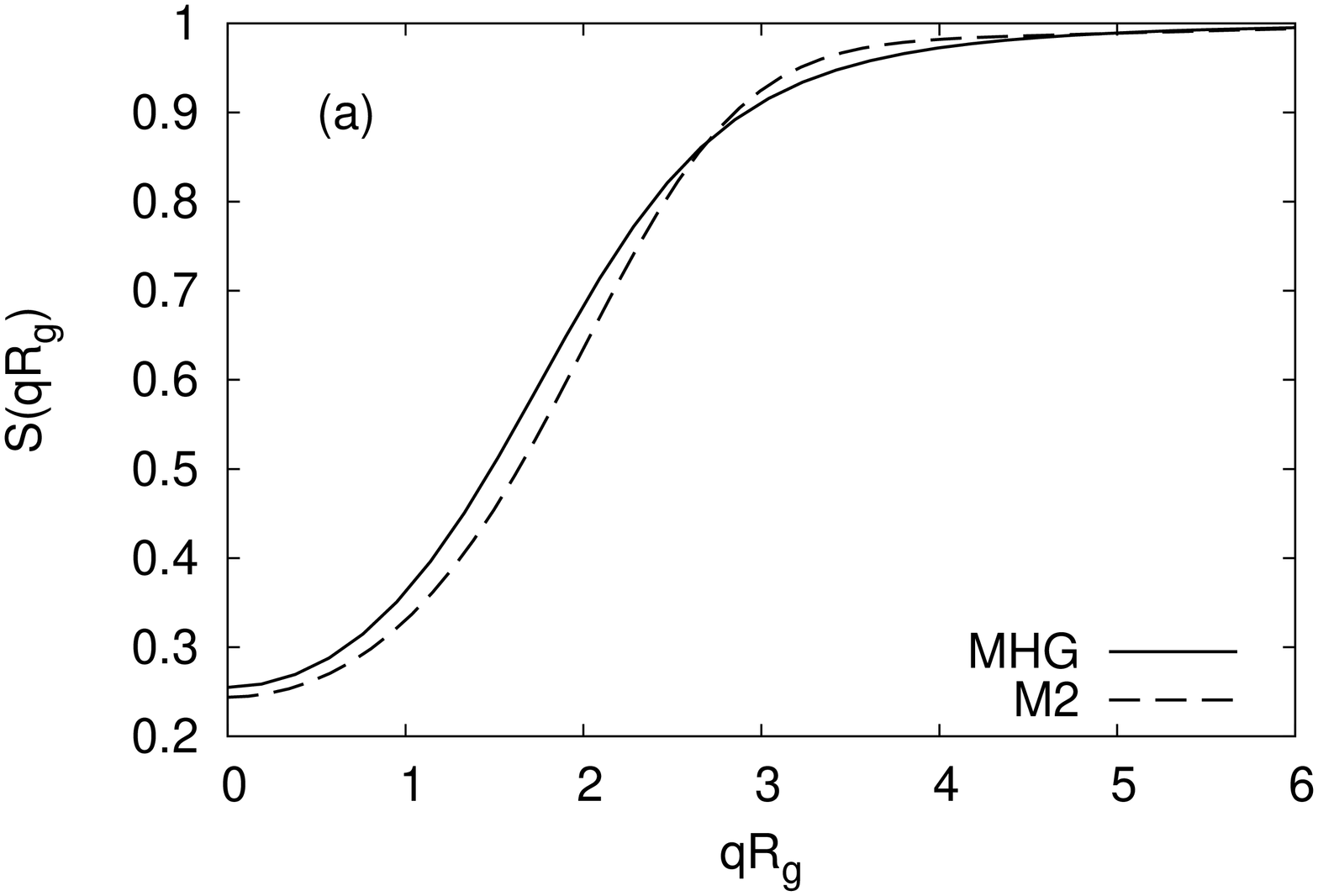,angle=-90,width=7truecm} \hspace{0.5truecm} &
\epsfig{file=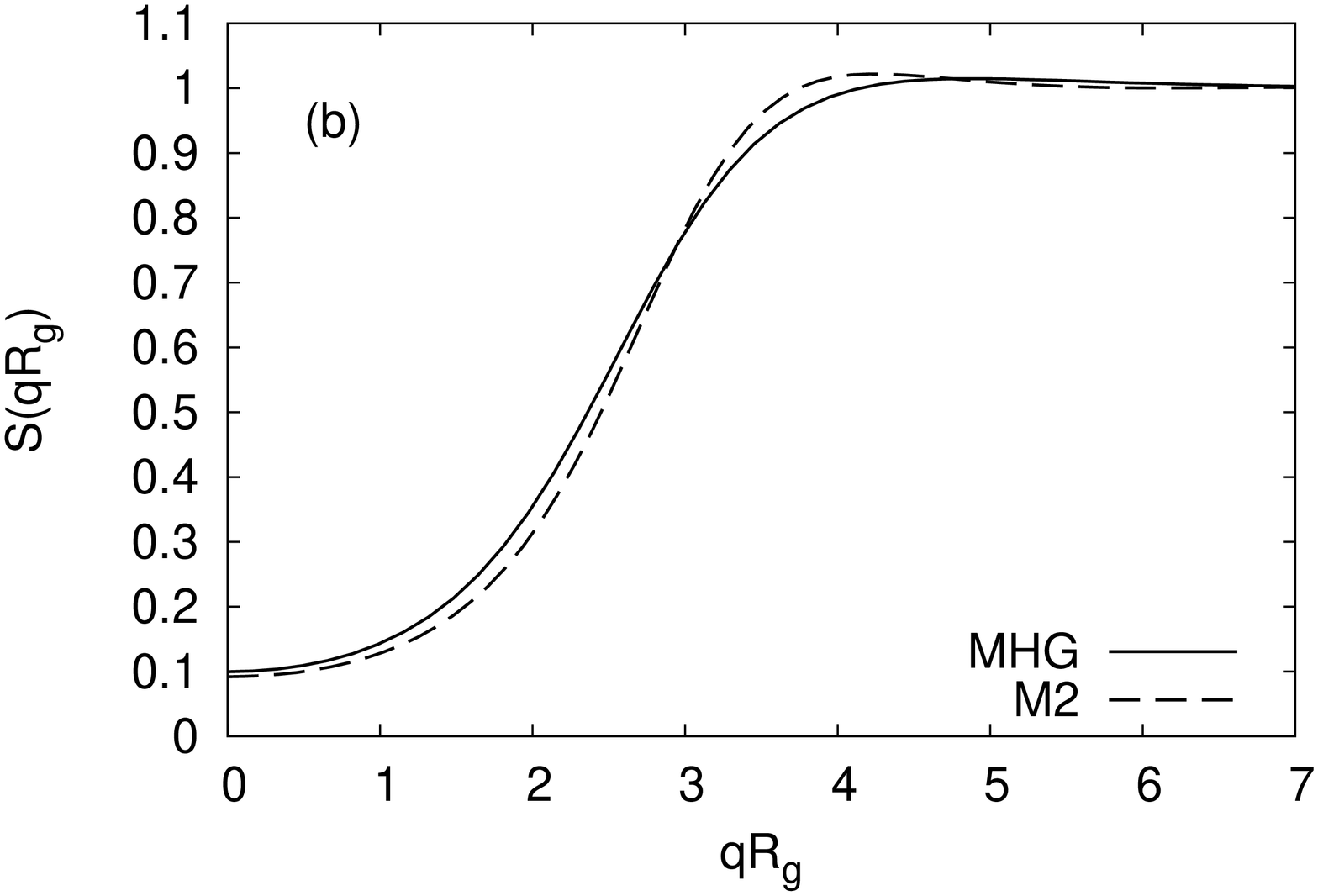,angle=-90,width=7truecm} \hspace{0.5truecm} \\
\epsfig{file=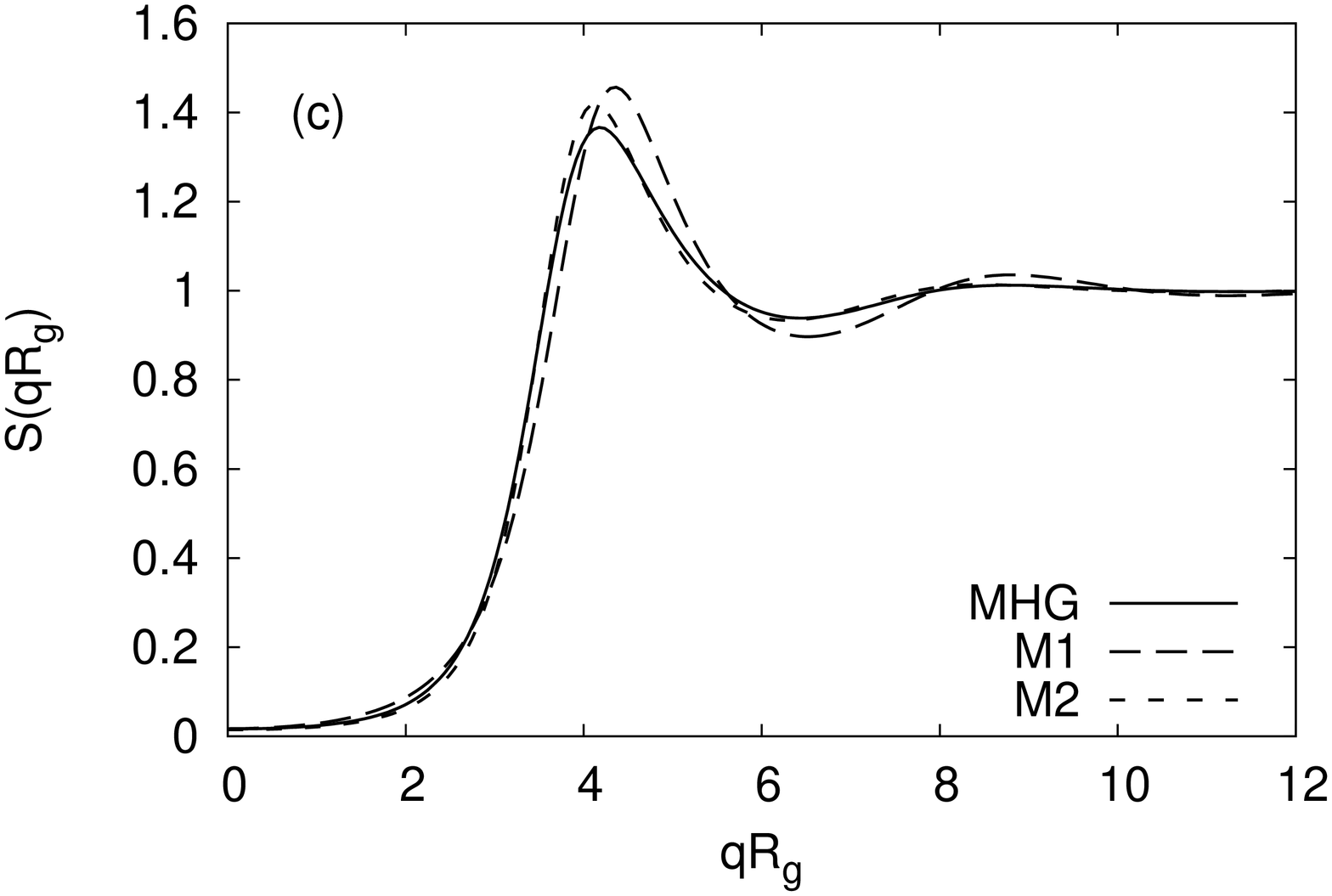,angle=-90,width=7truecm} \hspace{0.5truecm} &
\epsfig{file=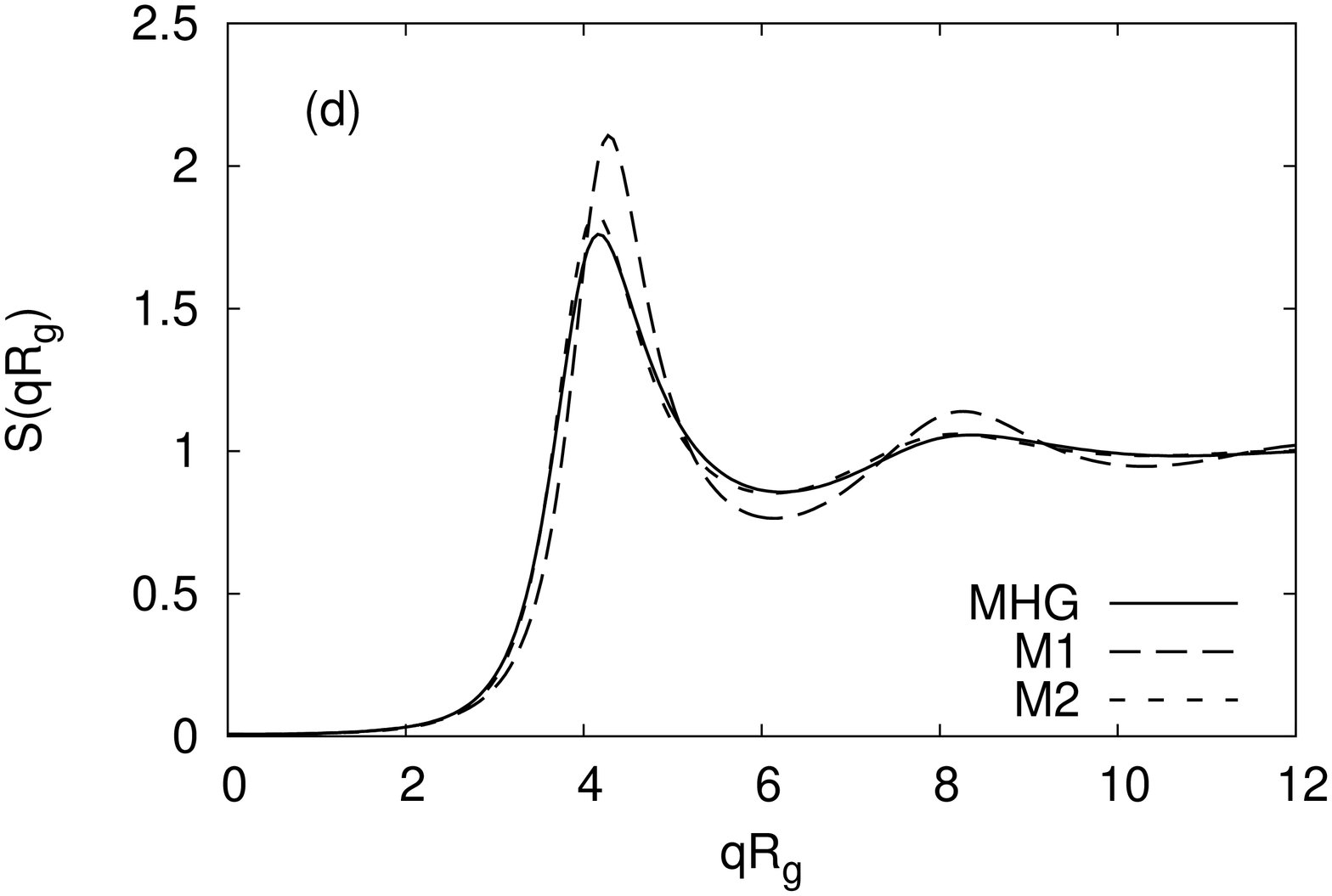,angle=-90,width=7truecm} \hspace{0.5truecm} \\
\end{tabular}
\caption{Structure factor as a function of $qR_g$ for $\Phi = 1$:
a) $f=2$ (models M2, MHG); 
b) $f = 5$ (models M2, MHG); 
c) $f = 18$ (models M1, M2, MHG); 
d) $f = 30$ (models M1, M2, MHG); 
For $f=18,30$ the results for model M2 are obtained by setting $\tau = 1$.
}
\label{figSq}
\end{figure*}

\subsection{Effective potentials with density-dependent corona diameter} 
\label{sec3.4}

Potential (\ref{potential-1}) 
was validated by comparing the theoretical predictions 
with experimental data for the total scattering intensity $I(q)$. 
In the dilute regime, $I(q)$ can be factorized \cite{foot-factoriz} as 
$I(q) \approx P(q) S(q)$, where $P(q)$ is the single-polymer form factor
and $S(q)$ is the intermolecular center-of-mass structure factor, which 
can be computed by using the three different CG models. 
In the CG model, $S(q)$ can be obtained as $S(q) = 1 + \rho \hat{h}(q)$,
where $\hat{h}(q)$ is the Fourier transform of the correlation function, 
which is obtained directly in the integral-equation calculation.

Results for $\Phi = 1$ are reported in 
Fig.~\ref{figSq} for $f=2,5,18,30$. 
The results obtained by using model M2 are quite similar to those 
obtained by using model MHG, both for $f=2,5$, which belong to the 
original validity range of the potential, and for $f=18,30$ (we take 
$\tau = 1$ in this case). On the other hand, results for model M1 
differ significantly.  The positions of the 
minima and maxima are the same for all potentials---not surprising since they 
are fixed by the dimensions of the polymer---but potential 
${\cal V}_1$ gives rise to significantly stronger oscillations for 
$f\ge 18$. 

The results presented here for model M1 are in apparent contradiction 
with the existing literature. Indeed, potential (\ref{potential-1}) 
has been extensively used to analyze experimental data, finding in all 
cases very good agreement. However, it should be noted that 
in all these comparisons a density-dependent corona diameter
is assumed. For instance, in the original paper\cite{LLWAJAR-98}
dealing with 18-arm polyisoprene in methylcyclohexane, $\sigma$ 
is fixed by $\sigma(\Phi) = 1.26 R_g(\Phi)$, where $R_g(\Phi)$
is the density-dependent radius of gyration. 
A careful study of the density dependence of the corona 
diameter is presented in Ref.~\onlinecite{Stellbrink-etal-00}
using 57-arm polybutadiene. They find  (see the inset of their Fig.~1)
that $\sigma(\Phi)$ decreases as $\Phi$ increases and that it behaves 
as $\Phi^{-1/8}$ for $\Phi\gtrsim 1$, in agreement
with the density scaling  predicted by the 
Daoud-Cotton model.\cite{DC-82}
Potential (\ref{potential-1}) 
has also been validated by using block-copolymer micelles with $f=63$. 
\cite{LSLWRZ-05}
However, as in previous cases, a density-dependent corona diameter is
assumed. 

\begin{figure}[tb]
\epsfig{file=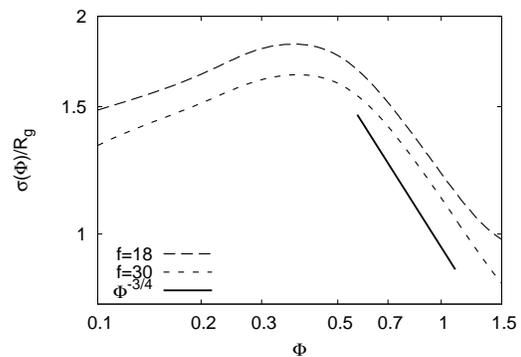,angle=-90,width=7truecm} \hspace{0.5truecm} 
\caption{Log-log plot of the density-dependent corona diameter 
$\sigma(\Phi)/R_g$ for $f=18$ 
and $f=30$. The solid line corresponds to a behavior $\Phi^{-3/4}$.
}
\label{fig:sigmaphi}
\end{figure}

\begin{figure*}[tb]
\begin{tabular}{cc}
\epsfig{file=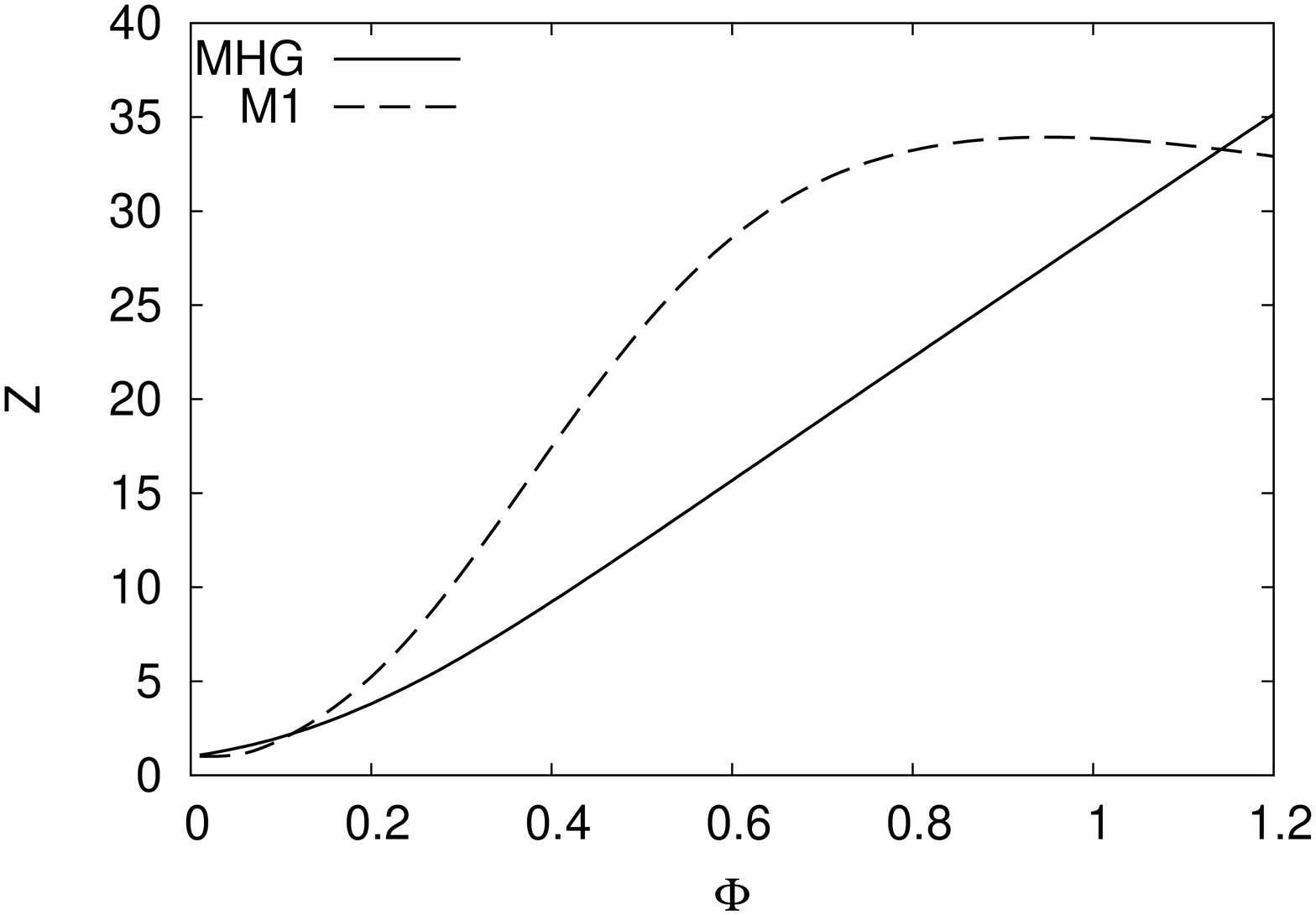,angle=-90,width=7truecm} \hspace{0.5truecm} &
\epsfig{file=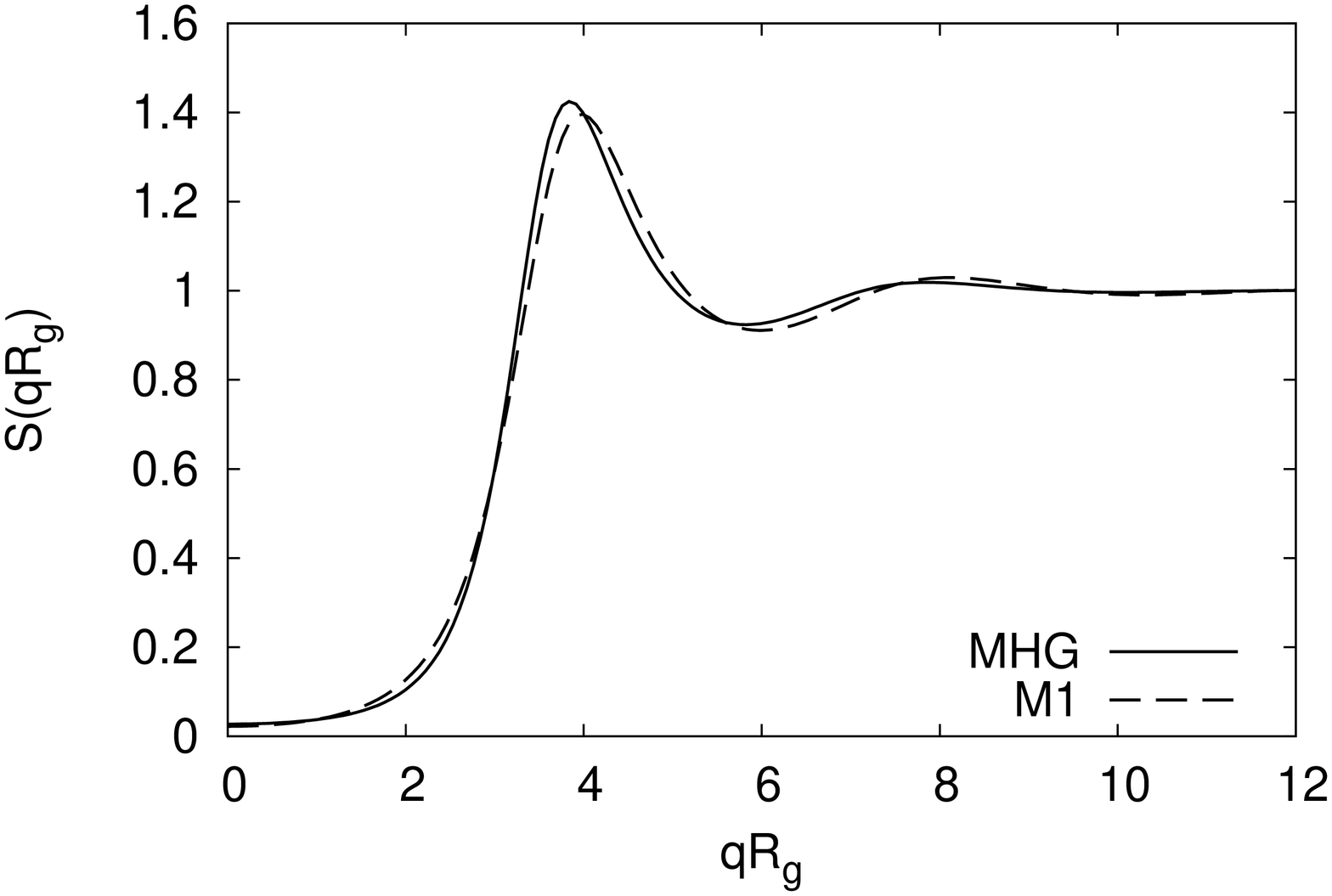,angle=-90,width=7truecm} \hspace{0.5truecm}
\\
\epsfig{file=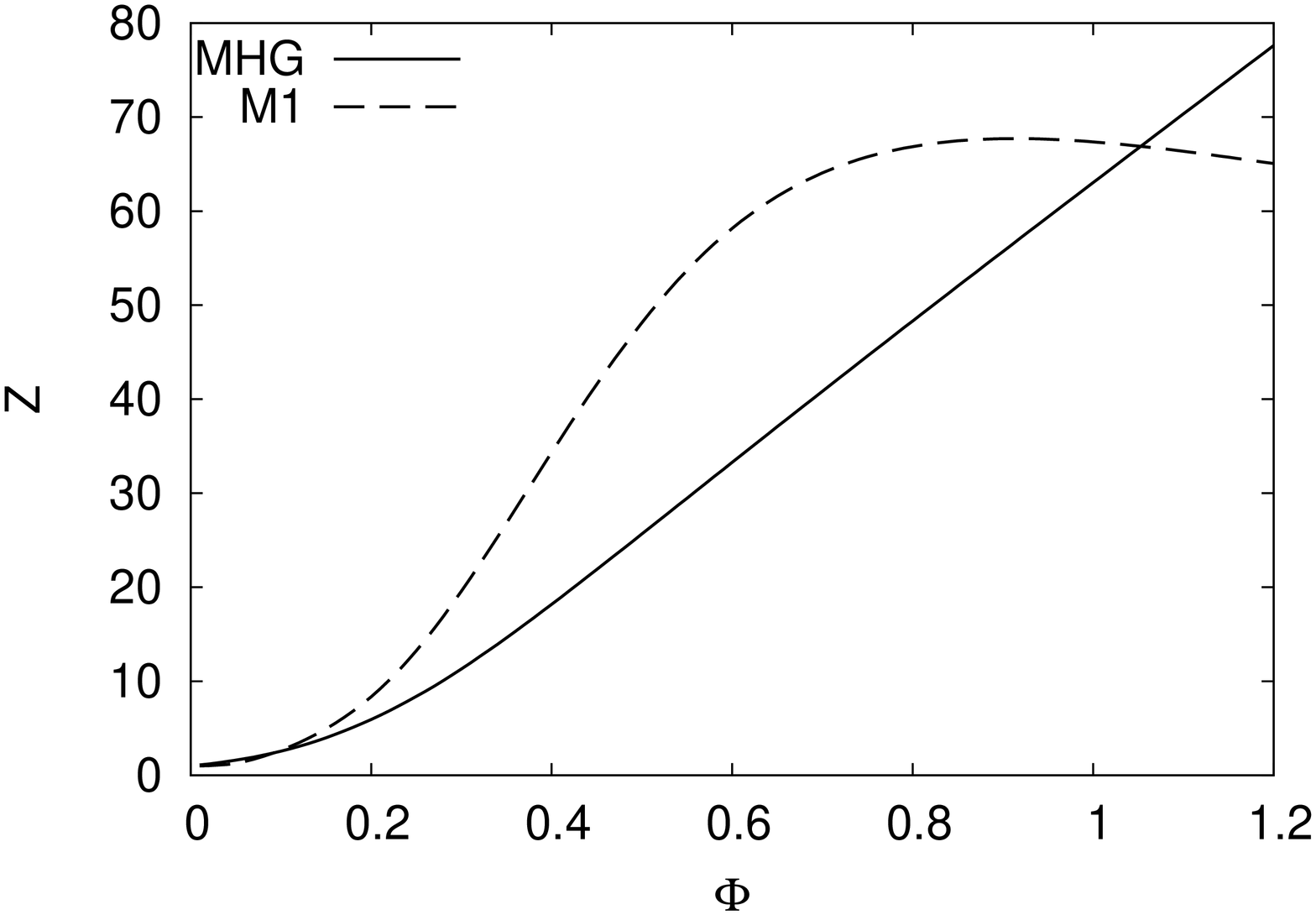,angle=-90,width=7truecm} \hspace{0.5truecm} &
\epsfig{file=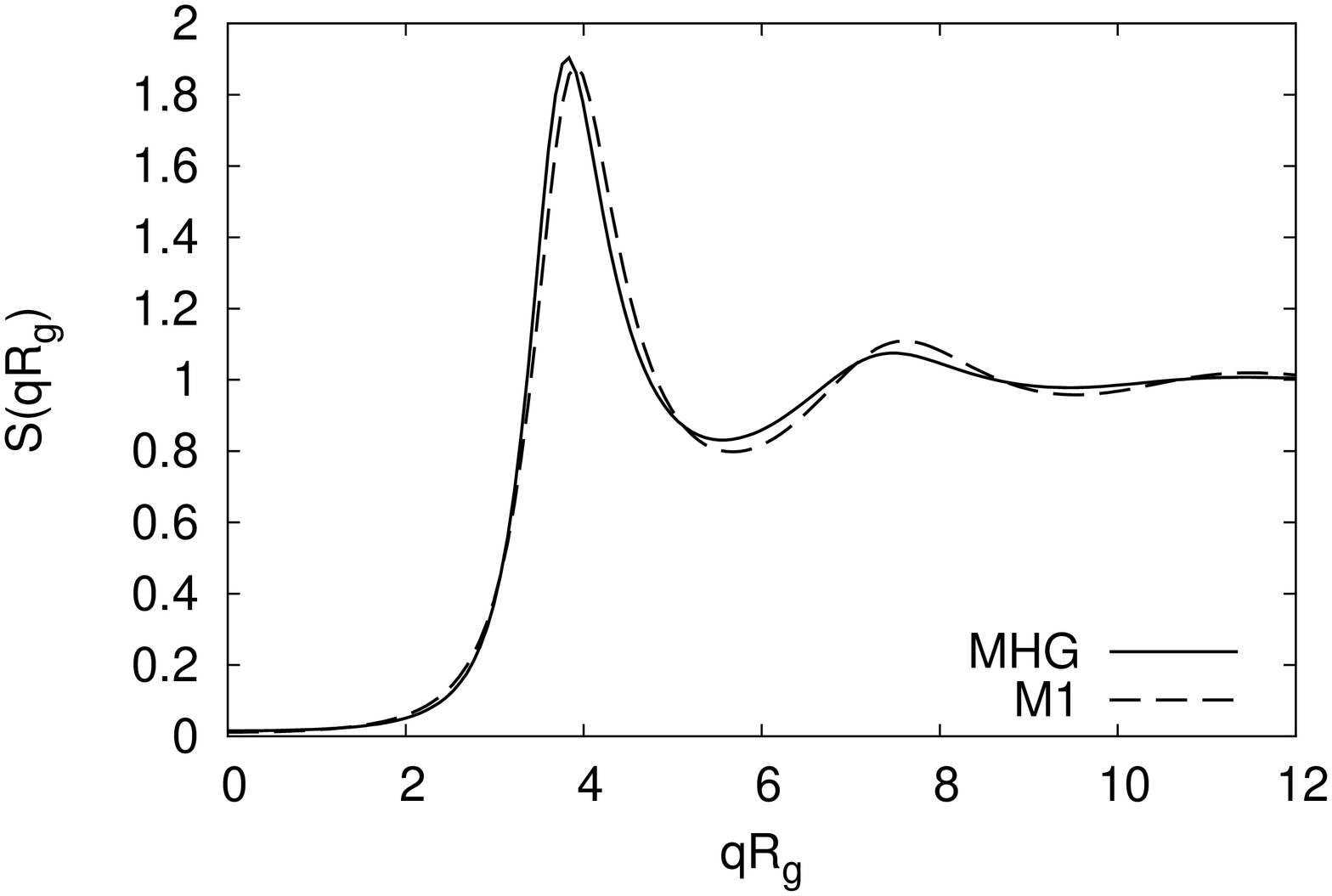,angle=-90,width=7truecm} \hspace{0.5truecm}
\end{tabular}
\caption{Compressibility factor $Z$ as a 
function of $\Phi$ (left)
and structure factors as a function of $qR_g$ for $\Phi = 1$ (right). 
Data for model M1 with the density-dependent
corona diameter reported in Fig.~\protect\ref{fig:sigmaphi}
and for model MHG.
Top: $f=18$; bottom: $f=30$. 
}
\label{fig:sigmavar-comparison}
\end{figure*}

To understand the role of a density-dependent corona diameter, we
repeat the same analysis as done in experimental work. We take 
the MHG structure factor as the reference one and determine 
a density-dependent $\sigma(\Phi)$ such that the M1 structure factor reproduces 
the MHG one at the same value of $\Phi$. Since the main difference 
between the MHG and M1 structure factors is the heigth of the peaks,
see  Fig.~\ref{figSq}, we fix $\sigma(\Phi)$ by requiring that 
the peak for the lowest value of $q$ has the same height in both models.
The results for $f=18$ and $f=30$ are reported in
Fig.~\ref{fig:sigmaphi}. In both cases, $\sigma(\Phi)$ increases
for $\Phi\lesssim 0.5$, then shows a maximum at a value which is 
significantly larger than that obtained by matching the 
second virial coefficients, and finally
decreases. The behavior looks similar to that observed in 
Ref.~\onlinecite{Stellbrink-etal-00} (see their Fig.~1). It is also
roughly consistent with the predictions of the 
Daoud-Cotton model, which predicts an intermediate regime  in which 
$\sigma(\Phi)\sim \Phi^{-3/4}$ before the onset of the large-$\Phi$ regime 
where $\sigma(\Phi)\sim \Phi^{-1/8}$. Model M1 with corona diameter
$\sigma(\Phi)$ well reproduces the MHG structure factor,
as can be seen from Fig.~\ref{fig:sigmavar-comparison}. 

The computation of the pressure in the presence of a density-dependent 
potential requires particular care.\cite{Louis:2002p2193} 
We use the HNC closure (thermodynamic consistency does not hold, hence we 
cannot implement the RY approach) and the compressibility route 
(the virial route does not provide the correct result,
see Ref.~\onlinecite{Louis:2002p2193}). 
The results are shown in Fig.~\ref{fig:sigmavar-comparison}.
It is evident that $Z$ for model M1 is quite different
from that of model MHG. But, even worse, the  M1 predicted compressibility 
factor shows an unphysical decrease for $\Phi\gtrsim 1$, a consequence of 
the quite rapid decrease of $\sigma(\Phi)$. Clearly, use of a density-dependent
corona diameter worsens the thermodynamic behavior of the model.

It is not surprising that $S(q)$ is well reproduced while large
differences are observed for $Z$. Indeed, the pair distribution function
and, therefore, also $S(q)$ are not very sensitive to the large-distance 
behavior of the potential: as discussed, for instance, in 
Ref.~\onlinecite{MullerPlathe-02}, visibly different potentials
may produce structures with essentially identical pair 
distribution functions. On the other hand, thermodynamic quantities
are very sensitive to the tail of the potential, hence differ
for models MHG and M1, even when a density-dependent corona diameter is 
used.

\section{Star polymer phase diagram} \label{sec4}

Since star polymers interpolate between linear chains, which only 
have a fluid phase for all densities, and hard colloids, which have 
a fluid-solid transition, star polymers 
are expected to behave in both ways depending on $f$.\cite{WPC-86,WP-86}
For $f < f_c$ one expects only a fluid phase, while for $f> f_c$ 
a fluid and a solid phase are expected. 
In Refs.~\onlinecite{WPC-86,WP-86}, $f_c$ was estimated to be of order 100. 
In Ref.~\onlinecite{WLL-99} a much more careful analysis was performed using 
model M1, finding $f_c \approx 34$. For $f_c < f \lesssim 60$ they 
found a small range of densities (note that 
$\Phi\approx 3.4\eta$, where $\eta$ is the volume fraction defined in 
Ref.~\onlinecite{WLL-99}), $1.5\lesssim \Phi\lesssim 2$, in which 
a solid bcc phase occurs, with reentrant melting as $\Phi$ increases. 
For larger values of $f$, the solid phase was more complex, with several
crystalline states appearing at different densities.
The presence of a solid phase for large values of $f$ was later confirmed
experimentally: Ref.~\onlinecite{LSLWRZ-05} observed a solid bcc
phase by using $f=67$ starlike block copolymer micelles.

Since the thermodynamical properties of model M1 are quite different from those
of model MHG, we wish now to check if and how the conclusions of 
Ref.~\onlinecite{WLL-99} change when potential (\ref{potential-HG}) is used. 
To determine the presence of a liquid-solid transition we use 
two different approximate methods. First, we use the Hansen-Verlet
criterion:\cite{HV-69,HS-73} the phase transition occurs when 
the maximum $S_{\rm max}$ of $S(q)$ exceeds 2.85. 
Second, we use the 
MHNC method\cite{RA-79,HansenMcDonald} and identify the phase 
transition by requiring 
$\eta_{HS} \approx 0.49$, where $\eta_{HS}$ is the packing 
fraction of the reference hard-sphere model. 

Since potential MHG has only been computed up to $f=35$ and the region
where the fluid-solid transition occurs corresponds to $f\gtrsim 30$, we 
have extended the MHG potential to $f=40$. For this purpose we need
to fix the five parameters that appear in Eq.~(\ref{potential-HG}).
Parameter $b_f$ is known,\cite{HG-04} $b_f = 94.6$. Then, we note that 
$a_f$ and $d_f$ have a tiny dependence on $f$, hence it should be 
safe to use a simple extrapolation of the values 
appropriate to $f=30$ and $35$: $a_f = 1.78$, $d_f=0.68$. 
Parameter $\tau_f$ is not precisely known and is approximately constant
for $f\gtrsim 10$. We take $\tau_f = 0.5$.
To fix $c_f$, we require the model to reproduce the second-virial
combination $A_2$: $A_2 \approx 41.96$ for $f=40$.\cite{Randisi-13} 
This gives $c_f = 84.2$. 

\begin{figure}[tb]
\vspace{0.5truecm}
\epsfig{file=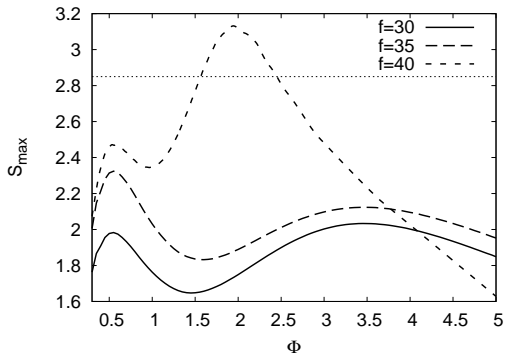,angle=-90,width=7truecm} \hspace{0.5truecm} 
\caption{Structure factor maximum $S_{\rm max} = \max S({q})$
as a function of $\Phi$ for $f=30,35,40$. We use the MHG model and the RY closure.
}
\label{fig:Hansen-Verlet}
\end{figure}

In Fig.~\ref{fig:Hansen-Verlet} we report the maximum $S_{\rm max}$ of 
$S(q)$ for $f=30,35,40$. For $\Phi < 5$, $S_{\rm max}$ shows a nonmonotonic
behavior with two maxima, one in the dilute region and one for $\Phi\gg 1$.
For $f=30$ and $35$ the maximum $S_{\rm max}$ is always smaller than 2.85,
hence no fluid-solid transition is expected. For $\Phi = 40$ instead, 
$S_{\rm max}$ is larger than 2.85 in the density range 
$1.6\lesssim \Phi\lesssim 2.4$ (the maximum corresponds to 
$\Phi \approx 1.94$ with $S_{\rm max} = 3.13$).
Hence, the Hansen-Verlet method allows us to infer that $35< f_c \lesssim 40$. 
Moreover, the 
solid phase should appear at values of $\Phi$ close to 2, 
i.e. in a range of densities that is similar to that reported in 
Ref.~\onlinecite{WLL-99}.

\begin{figure}[tb]
\vspace{0.5truecm}
\epsfig{file=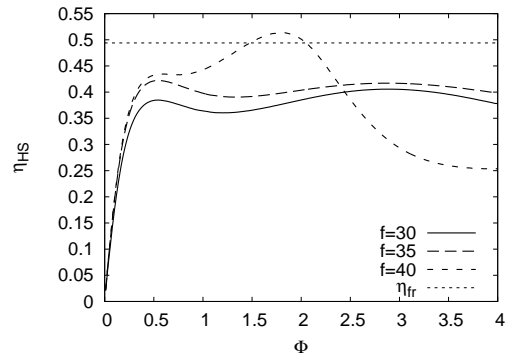,angle=-90,width=7truecm} \hspace{0.5truecm} 
\caption{
Effective hard-sphere packing fraction $\eta_{HS}$ determined by using 
the MHNC closure as a function of $\Phi$.
The horizontal line corresponds to the boundary of the fluid-solid
coexistence $\eta_{HS} = 0.49$.
}
\label{fig:MHNC-etaHS}
\end{figure}

The same analysis can be repeated by using the MHNC closure. 
In Fig.~\ref{fig:MHNC-etaHS} we report $\eta_{HS}$ as a function 
of $\Phi$ for $f=30,35,40$. For $f = 30,35$ the effective hard-sphere packing
fraction is always lower than 0.49, which gives the boundary of the 
fluid-solid coexistence line. For $f=40$, instead it reaches a 
maximum $\eta_{HS} = 0.51$ for $\Phi = 1.81$. Hence, the MHNC analysis
predicts crystallization in a small $\Phi$ interval that extends from 
$\Phi = 1.5$ to $\Phi = 2.0$. 

The results of the two analyses are fully consistent 
and allow us to conclude with confidence that $35 < f_c \lesssim 40$. 
We therefore confirm the conclusions 
of Ref.~\onlinecite{WLL-99}, although we predict the fluid-solid transition 
to occur for
slightly larger values of $f$ (Ref.~\onlinecite{WLL-99} predicted $f_c\approx 34$).

\section{Conclusions}  \label{sec5}

In this paper we investigate the thermodynamic behavior of three different
CG models appropriate to describe dilute star-polymer solutions. Model 
MHG uses the exact pair potential, model M1, which should be applied for 
$f > 10$, is inspired by the Daoud-Cotton model, while model M2 
is a phenomenological modification which shows a large-distance Gaussian
behavior and which is expected to be realistic for $f \lesssim 10$.
We find that model M2 provides a reasonable approximation to the thermodynamics
and to the structure, even for $f > 10$. On the other hand, model M1 
significantly underestimates the pressure and does not provide 
the correct structure factor. If one wishes to reconcile structural 
results for model M1 with those of 
model MHG, one might consider a density dependent corona diameter, 
as it is usually done in the analysis of the experimental data. 
However, while good agreement is obtained for the structure factor, 
model M1 with state-dependent interactions is completely inconsistent 
from a thermodynamic point of view: the compressibily factor shows an 
unphysical decrease as the density increases.

We investigate in detail the phase diagram of star polymers by using 
model MHG. We use both the Hansen-Verlet criterion 
and the MHNC approximation to estimate
the smallest value $f_c$ of the functionality $f$ for which a fluid-solid
transition occurs. Both analyses are consistent with $35 < f_c \lesssim 40$.
Our findings are in qualitative agreement with those of 
Ref.~\onlinecite{WLL-99}, which predicted $f_c\approx 34$ by using model M1,
and with experiments, which observed crystallization for $f\gtrsim 60$.
Our data are also consistent with the presence of reentrant melting 
for $f$ close to 40: the solid phase should be stable only in a 
small density interval 
centered around $\Phi\approx 1.8$-1.9.

\medskip

We thank Giuseppe D'Adamo for useful comments.

\end{document}